\documentstyle[12pt]{article}

\newcommand{\be}{\begin{equation}}
\newcommand{\ee}{\end{equation}}
\newcommand{\ba}{\begin{eqnarray}}
\newcommand{\ea}{\end{eqnarray}}

\begin{document}
\hoffset=-.4truein\voffset=-0.5truein 
\setlength{\textheight}{8.5 in}
\begin{center}
\hfill{LPTENS }

\vskip 0.6 in
{\large \bf  {Wilson's renormalization group :  a paradigmatic shift }}
\end{center}
\vskip .3 in
\begin{center}
 {\bf E. Br\'ezin} 

\end{center}
\vskip 5mm
\begin{center}
 Laboratoire de Physique
Th\'eorique, \\ { {\it
Unit\'e Mixte de Recherche 8549 du Centre National de la
Recherche Scientifique et de l'\'Ecole Normale Sup\'erieure.
 }} \\ brezin@lpt.ens.fr
\end{center}     
\vskip 3mm         
\begin{center}
{\bf Abstract}
\end{center}
\vskip 4mm
  A personal  and subjective recollection, concerning mainly Wilson's lectures delivered over the spring of 1972 at Princeton University (summary of a  talk  at Cornell University on November 16, 2013 at the occasion of the memorial Kenneth G. Wilson conference).  

\section{Why was it so difficult to accept Wilson's viewpoint?}
\setcounter{equation}{0}
\renewcommand{\theequation}{1.\arabic{equation}}
\vskip 4mm
In 1971-72 I was on leave from Saclay, employed by the theory group at the physics department of Princeton University. My background at the time was mainly field theory. Just before this I had been working with C. Itzykson on the use of  classical approximations in QED, dealing with problems such as pair creation by an oscillating electromagnetic field , but I was aware that the understanding  of the intriguing features of scaling laws and universality near a critical point was a difficult non-perturbative problem.
Ken Wilson was invited in the fall of 1971 to give a talk on his recently published  articles {\it{ Renormalization group and Kadanoff scaling picture \cite{W1} , Phase space cell analysis of critical behavior\cite{W2}}}
but Ken said that he could not explain his ideas in one talk.  The University \footnote{ thanks to David Gross as confirmed at the Cornell Wilson memorial conference} had the good idea of inviting him to talk as much as he needed and Ken ended up  giving 15 lectures in the spring term of 1972, which resulted in the well-known 1974 Physics Reports {\it{The renormalization group and the $\epsilon$-expansion}} by Wilson and Kogut (who had been taking notes throughout the lectures), one of the most influential articles of the last decades {\cite {KW}}. 

Wilson's ideas were so different from the standard views at the time, that they were not easy to accept. His style was also completely different : for him field theory was not a set of abstract axioms but  a tool to perform practical calculations even at the expense of oversimplified models whose relationship with reality was tenuous. Let us try to understand why Ken's ideas,  nowadays so inbred  in the training  of any young physicist that they seem nearly obvious, were so difficult to accept.  Let us first review the conventional wisdom at the time.
\begin{itemize}
\item Renormalization was a technique to remove perturbatively the ultra-violet singularities for those theories which were known to be renormalizable, i.e.  theories for which this removal could be done at the expense of a finite number of parameters. 
\item Why should one consider such theories? Presumably because they were the only ones for which practical calculations are conceivable. For instance in QED, the minimal replacement $p \to p-eA$ provides such a theory but gauge invariance alone would allow for more couplings, such as a spin current  coupled to the field strength, which would ruin the renormalizability.
\item As a result one obtained with renormalized QED a theory which, a priori, was potentially valid at all distances from astronomical scales down to vanishingly small ones. 
\item The renormalization group in its conventional formulation was applicable only to renormalizable theories ; it allowed one  to understand leading logarithms, at a given order in perturbation theory, from lower orders. 
\end{itemize}

So what is it which was surprising in Wilson's approach?
\begin{itemize}
\item  Critical points such as liquid-vapor, Curie points in magnets, Ising models, etc., are a priori classical problems in statistical mechanics. Why would a quantum field theory be relevant?
\item Why renormalization  in a theory in which there is a momentum cut-off $\Lambda = a^{-1}$ , in which $a$ is a  physical short distance such as a lattice spacing, or a typical interatomic distance, and there is no reason to let $a$ go to zero?
\item The momentum space reduction of the number of degrees of freedom that Wilson used introduced singularities in the Hamiltonian generated by the flow. It was not quite clear that the procedure could be systematized.
\item Irrelevance of most couplings was of course a fundamental piece of the theory but it looked a priori purely dimensional. For instance if one adds to a $g_4 \varphi^4$ theory in four dimensions a coupling $(g_6/\Lambda^2 )\varphi^6$ one obtains a flow equation for the dimensionless $g_6$ which concludes at its irrelevance as if the $1/\Lambda^2$ sufficed : 
$$ \Lambda \frac{\partial}{\partial \Lambda}g_6 - 2g_6 = \beta_6( g_4,g_6) $$
However the inclusion of the $g_6/\Lambda^2$ into a Feynman diagram with only $g_4$ vertices produces immediately an extra $\Lambda^2$ which cancels the previous one. So why is $g_6$ irrelevant?
\item Although it is now physically very clear, it was not easy to understand that the critical surface had codimension two for an ordinary critical point, in others words that there are two and only two relevant operators in the large space of allowed coupling parameters. 
\item Even more difficult to take was the statement that four-dimensional renormalizable theories  such as $\varphi^4$ or QED, i.e. non asymptotically free theories (an anachronistic name), if extended to all energy scales could only be free fields. How could one believe this, given the extraordinary agreement of QED with experiment? 
\end{itemize}
I must admit that my initial difficulties at grasping the views exposed in Ken's lectures led me, in a friendly collaboration with  David Wallace, to check their consistency. Amazed to see that the theory defeated our skepticism, we ended up working out with Ken the critical equation of state {\cite {BWW}}.

\section{Wilson's renormalization group}
\setcounter{equation}{0}
\renewcommand{\theequation}{2.\arabic{equation}}.
\vskip 2mm
It would take hundreds of pages to list all that we have learnt from Wilson's RG applied to critical exponents. The basic questions of scaling and universality were immediately made transparent,  all the critical indices could be expressed in terms of two  of them (plus a leading irrelevant operator for corrections to scaling, plus cross-over exponents). Wilson  insisted that RG was a computational tool even in the absence of small parameters, but he provided also new parameters.  Together with Michael Fisher, he introduced the famous epsilon-expansion\cite{WF} ; he was also the first to realize that critical properties could also be computed in a $1/n$ expansion (he pointed out that in a $(\vec\varphi^2)^2$ theory non bubble diagrams were subleading in the large $n$-limit), he contributed to Monte-Carlo RG, etc. I would like here to insist simply on the conceptual revolutions brought by  Wilson's RG: 
\begin{itemize}
\item  Field theory applied to the real world had to be considered as a long distance limit of yet unknown physics, even if the distances at which field theory is applied in particle physics go down to less than $10^{-6}$ femtometers. 
\item Renormalizability emerged as a consequence of considering this long distance limit, and not because the theory should be applicable down to vanishingly small distances.
\item Non asymptotically free theories such as $\varphi^4$, QED or the Weinberg-Salam model are merely {\it{effective}} theories. If one tried to  push them  down to vanishingly small distances they would be forced to be only free field theories. These theories carry their own limitation, without any indication on the underlying physics.
\end{itemize}

It is interesting to try to track down the maturation of his ideas. He introduced first the operator product expansion in an unpublished Cornell report of 1964 {\it{On products of quantum field operators at short distances}}. In his Nobel lectures Ken wrote {\it{I submitted the paper for publication; the referee suggested that the solution of the Thirring model might illustrate this expansion \footnote {it was reported at the Cornell conference quoted above that the referee was Arthur Wightman}. Unfortunately, when I checked out the Thirring model, I found that while indeed there was a short distance expansion for the Thirring model, my rules for how the coefficient functions behaved were all wrong, in the strong coupling domain. I put the preprint aside, awaiting resolution of the problem. }} He computed indeed the anomalous dimensions of the operators in the Thirring model (Operator-product expansions and anomalous dimensions in the Thirring model,\cite{W3}  : {\it{ It is shown that $\psi$ has dimension $1/2+\frac{ \lambda^2/4\pi^2}{1-\lambda^2/4\pi^2}$}}.) Clearly this had forced Ken to consider non-canonical dimensions of operators and the possibility that this could spoil the use of the OPE. Right after this he published his paper on . "Anomalous dimensions and breakdown of scale invariance in perturbation theory" \cite {W4}.  Other scientists at the time were also realizing that the Gell-Mann and Low RG could be interpreted as asymptotic restoration of scale invariance broken by the regularization of quantum fluctuations, in particular C. Callan ({\it{Broken scale invariance in scalar field theory}} \cite {Callan} )and K. Symanzik ({\it{Small distance behaviour analysis and Wilson expansions }}\cite{Symanzik}).  

The application of these ideas to critical phenomena was a natural illustration of Ken's views and he acknowledges in his 1971, Phys. Rev. B papers quoted here-above the role of Kadanoff's block spin picture\cite{Kad} and his conversations with Ben Widom and Michael Fisher at Cornell.  However Ken clearly was not very fond of the $\epsilon$ or $1/n$ expansions, although he had invented them. He regarded his RG as a tool that is applicable in the absence of any small parameter and thus the pseudo-small $\epsilon$ or $1/n$  were going back to directions that he didn't like.  Likewise he regarded the field theoretic method such as the Callan-Symanzik equations as a very special case of a general method. To prove his point the participants of the 1973 Carg{\`e}se summer school (which I co-organized)  remember that Ken arrived with a brilliant solution of the Kondo problem\cite{W5} (Renormalization group- critical phenomena and Kondo problem , Rev. Mod. Phys. 47, 773 (1975)) and his Nobel lectures he stated {\it{ [...] the solution of the Kondo problem is the first example where the full renormalization group (as the author conceives it) has been realized : the formal aspects of the fixed points, eigenoperators, and scaling laws will be blended with the practical aspect of numerical approximate calculations of effective interactions to give a quantitative solution [...] to a problem that previously had seemed hopeless.}} 

It is also at this Carg{\`e}se school  that he presented for the first time lattice gauge theories, which turned QCD calculations of mass spectra into a problem of statistical mechanics which remains as the main tool for understanding quantitatively the predictions of QCD.

\section{Quantitative aspects}
\setcounter{equation}{0}
\renewcommand{\theequation}{3.\arabic{equation}}

If there is no doubt forty years later that the RG has led to a complete and  extremely beautiful qualitative understanding of critical fluctuations,  succeeding to explain universality, scaling laws and many new phenomena such as the geometry of polymers, percolation, etc, the quantitative aspects are often less clear. Indeed the connection with the three dimensional physical world remains a difficult non-perturbative problem and various numerical strategies have to be devised. How systematic are some of these methods, such as real space RG, living with the size limitations of numerical simulations, has left room for analytical methods and I think that renormalized field theory has played a significant role. Indeed the theory deals with correlations extending to distances much larger than the lattice spacing  $ x\gg a$ ,  a correlation length $\xi \gg a$ and a finite ratio $x/\xi$. The effective renormalizable theory, the $a\to0$ limit, provides exactly all the correlation functions in this limit.  There are several strategies.

 For instance in the "minimal subtraction scheme" 
$$ \beta_d (g) = -(4-d) g +\beta_4(g)$$
$$ \gamma_d (g) =\gamma_4(g)$$
and all calculations in dimension $d$ follow  from computations in the critical (i.e. massless) theory in 4D. High order calculations in the $\epsilon$ or $1/n$ expansions follow from such techniques. However these expansions do not converge, no matter how small are $\epsilon$ or $1/n$. Using Lipatov's instanton method {\cite{Lipatov}} one can find the behavior of the perturbative series at order $k$  for large $k$ : for instance the successive terms of the $\epsilon$ expansion for the anomalous dimension of the field grow asymptotically like $\epsilon^k k! (-1/3)^k k^{7/2}$ (Ising like systems). This understanding led Zinn-Justin and co-workers {\cite {ZJ}},  using mathematical techniques such as Borel transform plus conformal maps, to accurate values of the critical exponents.

Parisi proposed to work directly  in three dimensions\cite {Parisi}, perturbation theory then makes sense only in the massive case, i.e. away from $T_c$. Again using the Lipatov type asymptotic estimate one can obtain at the end a good quantitative agreement with known values\cite{ZJ, BP}.

The point here is that it seems that one can transform the divergent perturbation series into seemingly  convergent algorithms.   Initially the impression was that one should better stop at some low order in the $\epsilon$ expansion if one did one want to run into absurdities and that was the best the theory could do. But now this says that the accuracy of the theory is not a priori limited : if comparison with experiment required better theoretical estimates, computing one more order in perturbation theory, although long and painful,  would still improve the theoretical accuracy. 

\vskip 2mm
\section{A crucial test of RG}
\setcounter{equation}{0}
\renewcommand{\theequation}{4.\arabic{equation}}
In 1969 in a remarkable pre-Wilson article, Larkin and Khmel'nitskii\cite{LK} made two interesting points
\begin{enumerate}
\item in 4D mean field theory is violated by powers of logarithms, and they computed explicitly those logs with a parquet approximation (i.e. sum of leading logs at each order) which is equivalent to a one-loop RG
\item if one considers an Ising-like magnet with strong dipolar forces, instead of the usual short-range exchange forces, then it is in 3D and not in 4D, that one should observe those logarithmic violations of mean field scaling.

\end{enumerate}
They predicted for instance that the specific heat, which in mean field theory remains finite at $T_c$ with a simple step when one crosses the critical temperature, would diverge instead as 
$$ C = A_{\pm} \vert \log \vert \frac {T-T_c}{T_c} \vert \vert^{1/3} $$
with
$$ \frac {A_+}{A_-} = \frac{1}{4}$$
Note that this is a prediction  which does not require any approximation scheme  ; it is a pure test of RG very similar in spirit to the logarithmic deviations to scaling in deep inelastic electron scattering  which provided the early tests of QCD{\cite{Gross}}. The experiment was performed in 1975 by G.Ahlers et al. {\cite{GA}} and their finding was : {\it{
Quantitative experimental results for the logarithmic corrections to the Landau specific heat are reported for the dipolar Ising ferromagnet $LiTbF_4$ near its Curie temperature $T_c=2.885 K$. The power of the leading logarithmic term is found to be $0.34\pm0.03$, and the corresponding amplitude ratio is $0.24\pm0.01$. These results are in agreement with the predicted values of $1/3$ and $1/4$, respectively.}}
\footnote{We realized with Zinn-Justin in 1975 {\cite{BZ}} that the correspondance between short range forces in 4D, a $\varphi^4$-theory with propagator $1/q^2$, and 3D with strong dipolar forces a $\varphi^4$-theory with propagator $1/q_{\perp}^2 + q_z^2/ q_{\perp}^2$, would not extend beyond one-loop.  A two-loop RG would give for the former $$C_{short range, d=4} = A_{\pm}  \vert \log \vert t \vert \vert^{1/3}  (1- \frac{25}{81} \frac{\vert \log \vert \log \vert t\vert \vert}{ \vert \log \vert t\vert\vert}) $$
and for the latter
$$C_{dipolar, d=3} = A_{\pm}  \vert \log \vert t \vert \vert^{1/3}  (1- \frac{1}{243}(41 +108 \log{\frac{4}{3}})  \frac{\vert \log \vert \log \vert t\vert \vert}{ \vert \log \vert t\vert\vert}) $$ 
A log-log is so slowly varying that this prediction has remained out of reach for experiments.}

\vskip 3mm
\section{A few remarks concerning RG and conformal invariance}
\setcounter{equation}{0}
\renewcommand{\theequation}{5.\arabic{equation}}
\vskip 2mm
The idea of asymptotic conformal invariance of strong interactions was present at an early stage in the work of Mack, Kastrup, Salam, Symanzik and others\cite{Mack2}. Wilson himself in his 1970 paper {\it{ Broken scale invariance and anomalous dimensions}}{\cite{Wilson}} refers to this asymptotic conformal symmetry. At the same period at the Landau Institute Polyakov {\cite{ Pol}} and Migdal \cite{Migdal} realized that the same conformal symmetry should hold near a critical point, and devised "bootstrap" equations to determine the theory. However this approach was temporarily abandoned for a number of reasons \begin{itemize}
\item The self-consistent bootstrap equation for the coupling constant did not contain any small parameter ... until the $\epsilon$-expansion was discovered.
\item This self-consistent equation was shown to be equivalent to the fixed point equation of RG\cite{Mack}
\item The full conformal group in generic dimension $D$ has $(D+1)(D+2)/2$ parameters, but if  the generators of invariance under rotations, translations, and dilatation are conserved, the remaining $D$ generators are conserved as well. Therefore there was not much interest to go beyond the standard RG. 
\end{itemize}
The point of view changed drastically for $D=2$ after the work of Belavin, Polyakov, Zamolodchikov \cite{BPZ} who pointed out that this was an infinite symmetry. The influence of this article on the last 30 years has been of course considerable. However it is only recently that the use of conformal symmetry plus crossing symmetry and unitarity was revived for three dimensional physics \cite{Slava}. Using this symmetry and bounds coming  from unitarity the authors determine an allowed domain in the plane $\Delta_{\sigma}, \Delta_{\epsilon}$, the dimensions of the spin and energy operators. For reasons which do not seem yet well understood the 3D Ising exponents lie on this boundary at a well-defined location. The authors also match their method  with the $\epsilon$-expansion and the agreement is impressive. \\

The theory is thus far from finished but without Wilson's RG which solved the problem of critical phenomena, and modified completely  the understanding of particle physics, we would still be in the stone age. 
\vskip 3mm


\end{document}